\documentclass[onecolumn,superscriptaddress,pra]{revtex4}%
\usepackage{graphics}

\usepackage{theorem}
\usepackage{enumerate}
\usepackage{amsmath}
\usepackage{amssymb}
\usepackage{graphicx}
\usepackage{epstopdf}
\usepackage{color}
\usepackage{euscript}
\usepackage{float}

\theorembodyfont{\rmfamily}
\newtheorem{Theorem}{\textit{Theorem} }
\newtheorem{Proposition}{Proposition}
\newtheorem{Definition}{\textit{Definition} }

\newcommand{\lv}{\left \vert}
\newcommand{\rv}{\right \vert}

\newcommand{\ra}{\right \rangle}
\newcommand{\ket}[1]{\lv #1 \ra}

\newcommand{\ketbra}[2]{\lv #1 \rangle \langle #2 \rv}
\newcommand{\tr}{{\rm tr}}

\begin{document}
\title{Diagonal quantum circuits: their computational power and applications}

\author{Yoshifumi Nakata}
\affiliation{
Institute for Theoretical Physics, the University of Hannover, Germany}
\email{yoshifumi.nakata@itp.uni-hannover.de}

\author{Mio Murao}
\affiliation{
Department of Physics, Graduate School of Science, University of Tokyo, Japan\\
Institute for Nano Quantum Information Electronics, University of Tokyo, Japan}

\begin{abstract}
Diagonal quantum circuits are quantum circuits comprising only diagonal gates in the computational basis.
In spite of a classical feature of diagonal quantum circuits in the sense of commutativity of all gates, 
their computational power is highly likely to outperform classical one and 
they are exploited for applications in quantum informational tasks. 
We review computational power of diagonal quantum circuits and their applications.
We focus on the computational power of instantaneous quantum polynomial-time (IQP) circuits, which are a special type of
diagonal quantum circuits.
We then review an approximate generation of random states as an application of diagonal quantum circuits,
where random states are an ensemble of pure states uniformly distributed in a Hilbert space.
We also present a thermalizing algorithm of classical Hamiltonians by using diagonal quantum circuits.
These applications are feasible to be experimentally implemented by current technology due to a simple and robust structure of diagonal gates.
\end{abstract}

\maketitle
\section{Introduction} \label{intro}
Quantum computation is believed to be more powerful than classical computation.
There are quantum algorithms that efficiently solve a problem for which no efficient classical algorithm is known so far,
such as the Shor's algorithm~\cite{S1997}, an algorithm for Pell's equation and the principal ideal problems~\cite{H2002,S2009}, and an algorithm for
approximate solutions of knot invariants~\cite{FLW2002,AJL2006,WY2008,AAEL2007}.
However, it has not been yet fully understood what makes quantum computation more powerful than classical one.
One approach to address this question is to investigate whether or not quantum computation described by circuits comprising a restricted class of quantum gates
still outperform classical computation.
From this perspective, it has been shown that, if the gates are restricted to those generating a little amount of entanglement~\cite{JL2003} or so-called matchgates acting on
nearest neighbor qubits~\cite{V2002,JM2008}, quantum computation is efficiently simulated by classical computers.

Quantum computation by diagonal quantum circuits in the computational basis with a separable initial state, which are often called 
{\it instantaneous quantum polynomial-time (IQP)} circuits, is also attracting much attention~\cite{SB2009,BJS2011,NV2012,JWAURB2013,FM2013}.
A study of diagonal quantum circuits is motivated by a fact that they are rather classical in the sense that all gates commute each other.
Hence, they are suited for investigating the boarder of quantum and classical computational power.
Since noncommutativity of operators is one of the significant features of classical theory,
one may expect that diagonal quantum circuits would not outperform classical computers.
This natural expectation is however likely to be incorrect.
In Ref.~\cite{BJS2011}, it has been shown that IQP circuits are not classically simulatable under the assumption that the polynomial hierarchy (PH) does not collapse at the third level,
which is a highly plausible assumption in computational complexity theory.
Such computational power of IQP circuits is intuitively related to
a fact that a quantum circuit diagonal in the computational basis 
can generate highly entangled state by choosing an appropriate separable initial state.
In fact, almost all states generated by IQP circuits are extremely entangled~\cite{NTM2012}.
These results clarify the distinction between quantum and classical computational power,
and give an insight that quantum theory is not necessarily reduced to classical one by imposing only commutativity of operators.

Diagonal quantum circuits also have a practical importance
since they are experimentally much simpler to realize and less sensitive to decoherence than non-diagonal quantum circuits.
Diagonal gates are fault-tolerantly realizable even by current technology, for instance, in superconducting and semiconducting systems~\cite{ABDPST2009}.
Moreover, the commutativity of all gates in a diagonal quantum circuit allows us to implement the circuit by a single time-independent commuting Hamiltonian.
In such an implementation by a Hamiltonian dynamics, a control of the order of interactions is not necessary since all interactions can be simultaneously applied due to their commutativity,
so that it significantly reduces practical time to perform computation and makes the implementation more robust.
For these reasons, it is worth studying what diagonal quantum circuits can perform beyond classical information processing,
which directly provides experimentally realizable quantum tasks by currently available technology.

In spite of their computational power and practical merits of diagonal quantum circuits, little is known about concrete applications of diagonal quantum circuits so far.
One of the applications of quantum diagonal circuits is a generation of {\it random states}, which are an ensemble of pure states uniformly distributed 
in a Hilbert space with respect to the unitarily invariant measure.
Random states have many utilities in a wide range of applications in quantum information processing such as
quantum communicational tasks~\cite{L1997}, efficient measurements~\cite{RBSC2004}, an algorithmic use~\cite{RRS2005,S2006} and an estimation of gate fidelities~\cite{DCEL2009}.
However, generating random states requires exponential resource since the number of parameters in random states scales exponentially with the number of qubits.
Hence, efficient generations of approximate random states, called a {\it state $t$-design}, have been intensely studied~\cite{DCEL2009,EWSLC2003,DLT2002,ODP2007,DOP2007,Z2008,HL2009,DJ2011,HL2009TPE,BHH2012,CHMPS2013}.
A $t$-design of an ensemble is another ensemble that simulates up to the $t$th-order statistical moments of the original one~\cite{RBSC2004,AE2007}.
A state $t$-design can be approximately but efficiently generated by a quantum circuit~\cite{BHH2012} and can be used instead of random states in
many applications~\cite{L2009}.
In Refs.~\cite{NM2013,NKM2013}, it was shown that non-diagonal gates are not necessary for generating a state design
by presenting a way of approximately generating a state $t$-design by diagonal quantum circuits. Although the degree of approximation is small but constant and 
cannot be improved by applying additional diagonal gates in general,
it was also shown that a state $2$-design can be exactly generated by combining a diagonal quantum circuit with a classical random procedure~\cite{NM2013}.
These results are practically useful since they provide a concrete application of diagonal quantum circuits in
quantum informational tasks exploiting a state $t$-design.
This allows us to experimentally demonstrate quantum advantages.

This paper aims to review the results on diagonal quantum circuits in terms of their computational power and applications.
It is organized as follows.
The computational power of diagonal quantum circuits is summarized in Sect.~\ref{sec:CompPower}.
The rest of the paper is devoted to overview the results obtained in our previous papers~\cite{NM2013,NKM2013} about applications of diagonal quantum circuits,
which are given in Sect.~\ref{sec:App}.
We summarize the paper with concluding remarks in Sect.~\ref{sec:Sum}.

Before leaving the introduction, we explain our notation in this paper.
We consider an $n$-qubit system, where its Hilbert space is denoted by $\mathcal{H}=(\mathbb{C}^2)^{\otimes n}$.
The eigenstates of the Pauli-$Z$ ($X$) operator are denoted by $\ket{0}$ and $\ket{1}$ ($\ket{+}$ and $\ket{-}$), which corresponds to the eigenvalues $1$ and $-1$, respectively.
They are related by $\ket{\pm} = (\ket{0} \pm \ket{1})/\sqrt{2}$.
We define the computational basis by tensor products of $\ket{i}$ ($i=0,1$), and denote it by $\{ \ket{\bar{m}} \}_{m=0,\cdots, 2^n-1}$, where $\bar{m}$ is a binary representation of $m$. 
Throughout the paper, we study diagonal quantum circuits in the computational basis, so that we do not refer to the basis in the following.

\section{Computational power of diagonal quantum circuits} \label{sec:CompPower}

We review results on the computational power of diagonal quantum circuits mainly obtained in Refs.~\cite{BJS2011,FM2013}.
In Sect.~\ref{ssec:ComClass}, we provide a very brief introduction of some complexity classes, which will be used in the following subsections.
The results of computational complexity on IQP circuits are summarized in Sect.~\ref{ssec:IQP}.

\subsection{Computational complexity classes} \label{ssec:ComClass}

We briefly explain computational complexity classes for decision problems, which are the problems with yes-or-no answers.
We overview only the complexity classes related to an investigation of IQP circuits. For more details, see, e.g., Refs.~\cite{AB2009,P1994}.

We first explain complexity classes called {\it polynomial-time (P)} and {\it nondeterministic polynomial-time (NP)}.
The class P is a class of the problems solvable in polynomial time by classical computers.
The class NP is problems with the following properties:
if the answer is yes, there exists a proof of polynomial length to confirm this fact that can be verified in P,
and if the answer is no, all proofs purported to that the answer is yes are rejected in P.
The class P is clearly included in NP, but it has not been shown whether the inclusion is strict or not.
From the complexity classes P and NP, a {\it polynomial hierarchy (PH)} is defined based on an idea of oracle machines.
A class $\Delta_k$ ($k=1,2,\cdots$) in the hierarchy are recursively defined by $\Delta_1=P$ and $\Delta_{k+1}=P^{N\Delta_k}$,
where $N\Delta_k$ is the non-deterministic class associated with $\Delta_k$, e.g., $N\Delta_1=NP$.
The ${\rm P}^{N\Delta_k}$ is a set of problems that are in $P$ if an oracle belonging to a complexity class $N\Delta_k$ is allowed to use.
Since one use of an oracle is counted as one step, ${\rm P}^{N\Delta_k}$ is expected to to be much harder than  $N\Delta_k$ although it has not been proven.
A complexity class PH is then defined by the union of all classes $\Delta_k$ (see e.g.~\cite{AB2009}).

In contrast to a deterministic feature of P, NP and PH in the sense that the computation is deterministically performed,
there are probabilistic complexity classes.
In particular, two classes {\it bounded-error probabilistic polynomial-time (BPP)} and {\it probabilistic polynomial-time (PP)} are important for investigating IQP circuits.
The problems in BPP can be solved probabilistically in polynomial time, where the probability of obtaining a correct answer should be greater than or equal to $2/3$.
If the probability of obtaining a correct answer is relaxed to be strictly greater than $1/2$, the class of the problems is called PP.
The distinction between BPP and PP is whether or not the lower bound of the probability that the algorithm provides a correct answer,
which should be strictly greater than $1/2$, can depend on the input size.
For BPP, the probability $2/3$ can be replaced with an arbitrary $p>1/2$ as long as it is constant, which enables us to amplify the probability arbitrary close to 1 by running the algorithm polynomially many times.
On the other hand, the lower bound of the probability is possibly dependent on the input size for PP, e.g., $p>1/2 + 1/2^n$. Hence, BPP is a subclass of PP. 

\begin{figure}[tb]
\centering
\includegraphics[width=60mm, clip]{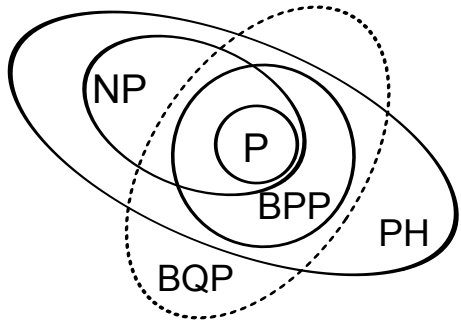}
\caption{A diagram for the relations between complexity classes P, NP, PH, BPP, and BQP. Note that it is not clear whether or not the inclusions are strict.}
\label{Fig:CC}
\end{figure}

For quantum computation, we explain only {\it bounded-error quantum polynomial-time (BQP)}~\cite{BV1993}, which is a quantum version of BPP. 
Formally, BQP is a class of problems solvable by a quantum computer in polynomial time, where the probability to obtain a correct answer is greater than or equal to $2/3$.
The class BQP includes BPP and P, and is included in PP~\cite{ADH1997}, so that P $\subseteq$ BPP $\subseteq$ BQP $\subseteq$ PP.
Although the relation between BQP and NP is not exactly known, 
there is an evidence that BQP is not included in PH~\cite{A2009}, implying that BQP is likely to be not included in NP. See also Fig.~\ref{Fig:CC}

Finally, we introduce computational classes with a post-selection.
For every probabilistic computational class $A$, 
its post-selected version can be defined by allowing a post-selection, which is simply denoted by post-A,
e.g., post-BPP, post-PP, and post-BQP.
Although a post-selection may not be realistic,
it helps an investigation of relations between different complexity classes for which a post-selection is not allowed.

\subsection{IQP circuits and their computational power} \label{ssec:IQP}

Instantaneous quantum polynomial-time (IQP) circuits are quantum circuits comprising only diagonal gates in the computational basis with a separable initial state and 
measurement in a separable basis.
In spite of the commutativity of diagonal gates, they have quantum features 
if the initial state and the measurement basis are appropriately chosen.
The definition of IQP circuits is given by the following.

\begin{Definition}[{\bf IQP circuits~\cite{SB2009,BJS2011}}]
{\it An IQP circuit for $n$ qubits is a quantum circuit with the following structure:
each gate in the circuit is diagonal in the Pauli-$Z$ basis, the input state is $\ket{+}^{\otimes n}$,
and the output is the result of a measurement in the Pauli-$X$ basis on a specified set of output qubits.}
\end{Definition}

A complexity class related to IQP circuits is simply denoted by IQP.
Its classical simulatability has been studied in Refs.~\cite{BJS2011,FM2013},
where the simulatability of quantum circuits is defined in two ways, a {\it strong} and {\it weak} simulation~\cite{BJS2011}.

\begin{Definition}[{\bf Simulatability of quantum circuits}] \label{Def:SimQC}
{\it Let $\mathcal{C}$ and $P(\mathcal{C})$ be a quantum circuit of a uniform family and the probability distribution obtained by a 
measurement in a separable basis, respectively.
A circuit family $\mathcal{C}$ is strongly simulatable if the output probability distribution $P(\mathcal{C})$ can be computed in polynomial time.
A circuit family $\mathcal{C}$ is weakly simulatable if there exists a method to sample the output according to
the probability distribution $P(\mathcal{C})$ in polynomial time.}
\end{Definition}

Although it is clear that strong simulatability implies weak one, their difference is large
since there exist quantum circuit families that are classically weakly simulatable but its exact strong simulation is in a complexity class $\sharp P$-complete~\cite{V2010}, which is strongly believed to be a much harder than P.
As the exact simulation of quantum circuits is too restrictive, an approximate simulation is also introduced in terms of a multiplicative error
for a weak simulation of quantum circuits, e.g., in Ref.~\cite{BJS2011}.
A circuit family $\mathcal{C}$ is weakly simulatable with a multiplicative error $c \geq 1$ 
if there is a sampling method according to the probability distribution $\tilde{P}(\mathcal{C})$, where
\begin{equation}
\frac{1}{c} P(\mathcal{C}) \leq \tilde{P}(\mathcal{C}) \leq c P(\mathcal{C}).
\end{equation}
The following theorem is about classical simulatability of IQP circuits.

\begin{Theorem}[{\bf Hardness of classical simulations of IQP circuits by Bremner et al.~\cite{BJS2011}}] \label{Thm:IQPhard}
{\it If the output probability distributions generated by IQP circuits could be classically weakly simulated to within multiplicative error
$1 \leq c < \sqrt{2}$ then PH $= \Delta_3$. This is the case even for  IQP circuits with two-qubit gates.}
\end{Theorem}

The proof of Theorem~\ref{Thm:IQPhard} relies on an investigation of a post-selected version of IQP circuits.
It was first shown that if a post-selection is allowed, computational power of IQP circuits is equal to that of BQP.
This is interesting in itself since it implies that a post-selection closes the gap between diagonal and non-diagonal circuits.
The computational power of post-BQP is also known to be equal to PP~\cite{A2005}, resulting in a relation that
\begin{equation}
{\rm post {\text -} IQP} = {\rm PP}. \label{Eq:ccc}
\end{equation}
Moreover, under the assumption that IQP circuits are weakly classically simulatable as stated in Theorem~\ref{Thm:IQPhard},
it was shown that post-IQP $\subseteq$ post-BPP.
This relation and Eq.~\eqref{Eq:ccc} imply, together with a fact post-BPP $\subseteq$ post-BQP $=$ PP~\cite{A2005}, that post-BPP $=$ PP.
By using relations that PH $\subseteq$ P$^{\rm PP}$~\cite{T1991} and P$^{\rm post {\text -} BPP} \subseteq \Delta_3$~\cite{HHT1997},
it is obtained that PH $\subseteq \Delta_3$, meaning a collapse of the polynomial hierarchy at its third level (see Ref.~\cite{BJS2011} for the full detail). 

Since it is highly implausible that PH $= \Delta_3$,
Theorem~\ref{Thm:IQPhard} implies that IQP circuits are highly unlikely to be weakly simulatable by classical computers.
One intuitive understanding of this computational power of IQP circuits
is obtained by studying the properties of states that appear during the computation by IQP circuits.
Although IQP circuits are capable to use only a restricted ensemble of states due to the commutativity of all gates and a fixed separable initial state,
it was shown that such an ensemble of states covers the whole Hilbert space fairly uniformly in the sense that 
the ensemble is hard to be distinguished from the uniformly distributed states in the Hilbert space~\cite{NKM2013}. 
Consequently, a state generated by an IQP circuit is typically highly entangled~\cite{NTM2012}.
These results are shown in the formulation of a $t$-design of random states, which we will explain in more detail in Sects.~\ref{ssec:RMRS} and~\ref{sssec:Approx}.
A fact that IQP circuits can exploit such a uniformly distributied states in a Hilbert space indicates that the outputs of the circuits 
should typically have quantum features in spite of the commutativity of all gates,
and is expected to result in the computational power exceeding the classical one.

Although Theorem~\ref{Thm:IQPhard} shows that there is no universal classical method to simulate an output distribution of {\it any} IQP circuits,
it is also interesting to address a question {\it what types of IQP circuits are classically simulatable}.
This question is partially answered by mapping the output probability distribution to the partition function of the Ising models with imaginary coupling constants
on an associated graph~\cite{FM2013}.
The following Theorem is obtained by investigating whether or not the partition function of the mapped Ising model is exactly computed:

\begin{Theorem}[{\bf Classically simulatable subclass of IQP circuit by Fujii et al.~\cite{FM2013}}] \label{Thm:CSIQP}
{\it If IQP circuits are sufficiently sparse,
or if IQP circuits contain only two-qubit gates of the form $\exp[ i Z_l \otimes Z_m ]$ acting on nearest-neighbor qubits $(l,m)$ on a two-dimensional graph,
then the IQP circuits are classically simulatable in a strong sense.}
\end{Theorem}
The condition of the sparse property in Theorem~\ref{Thm:CSIQP} is given by an associated graph of IQP circuits, which is obtained by identifying qubits and diagonal gates in the circuit with vertices and hyperedges on a graph. If the associated graph is an {\it independent and full rank bipartite (IFRB)} graph, the sparse condition is satisfied
(see Ref.~\cite{FM2013} for details).

Theorem~\ref{Thm;CSIQP} implies that there exist IQP circuits that are easy to classically simulate. 
Although only two specific cases were found in Ref.~\cite{FM2013},
the method developed in the paper, mapping an output probability distribution of IQP circuits to the partition function of the Ising models, 
works in general and will provide a good tool for studying classical simulatability of IQP circuits.\\

Theorems~\ref{Thm:IQPhard} and~\ref{Thm:CSIQP} imply that IQP circuits show both classical and quantum computational power depending on the detailed structure of the circuits and that their computational power is on the border of classical and quantum ones.
Thus, IQP circuits will provide a good framework to address a question what property of quantum circuits distinguishes quantum from classical in terms of computational power.
Although this question can be addressed by investigating non-diagonal quantum circuits, IQP circuits are probably more suitable since they have 
a much simpler structure than non-diagonal ones.

\section{Applications of diagonal quantum circuits} \label{sec:App}

In this section, we provide two applications of diagonal quantum circuits.
Since the applications are closely related to random unitary matrices and random states, we explain them in Sect.~\ref{sec:DUM}.
We also introduce random diagonal-unitary matrices and phase-random states, which are a restricted version of random unitary matrices and random states, respectively.
We then show how to implement a $t$-design of random diagonal-unitary matrices by diagonal quantum circuits in Sect.~\ref{sec:DUbyIQP}.
This guarantees an efficient realization of the applications of random diagonal-unitary matrices by diagonal quantum circuits.
In Sects~\ref{ssec:Gt} and~\ref{ssec:Therm}, we provide two applications of random diagonal-unitary matrices, generating a state $t$-design~\cite{NKM2013,NM2013} and a thermalizing algorithm for classical Hamiltonian.

\subsection{Random matrices, random states and $t$-designs} \label{sec:DUM}

We overview definitions and applications of random unitary matrices, random states and their $t$-designs.
In Sect.~\ref{ssec:RMRS}, we explain random unitary matrices and random states and 
briefly summarize their properties and applications.
A study of $t$-designs of random unitary matrices and random states is summarized in Sect.~\ref{ssec:t}.

\subsubsection{Random unitary matrices and random states} \label{ssec:RMRS}

Random unitary matrices are originally introduced in the field of random matrix theory~\cite{M1990} and defined by an ensemble of unitary matrices uniformly distributed in a 
unitary group with respect to the Haar measure.
Similarly, random states are an ensemble of pure states distributed uniformly in a Hilbert space,
which are obtained by applying random unitary matrices on a fixed initial state.
Their definitions are given by the following.

\begin{Definition}[{\bf Random unitary matrices~\cite{M1990} and Random states}]
{\it Let $\mathcal{U}(d)$ be the unitary group of degree $d$. Random unitary matrices $\mathcal{U}_{\rm Haar}$ are the ensemble of unitary matrices uniformly distributed in $\mathcal{U}(d)$ with respect to the Haar measure $d\mu_{\rm Haar}$.
Random states are the ensemble of states uniformly distributed in a Hilbert space, which are given by
$\{ U_{\mu} \ket{\Psi} \}_{U_\mu \in \mathcal{U}_{\rm Haar}}$ for any fixed state $\ket{\Psi}$ in a Hilbert space with dimension $d$.}
\end{Definition}

Note that the distribution of $\{ U_{\mu} \ket{\Psi} \}_{U_\mu \in \mathcal{U}_{\rm Haar}}$ is independent of 
the choice of $\ket{\Psi}$ due to the unitary invariance of the Haar measure, i.e., $d\mu_{\rm Haar}(U)=d\mu_{\rm Haar}(UV)=d\mu_{\rm Haar}(V)$ for any $U,V \in \mathcal{U}(d)$.
The unitarily invariant property of random unitary matrices and random states leads to many uses of them in the field of quantum information. 
Random unitary matrices are used for cryptographic use~\cite{BR2003,AMTW2000,AS2004,DN2006,HLSW2004,Au2009}, quantum communication tasks~\cite{L2000,HHL2004}, a quantum data hiding~\cite{DLT2002,TDL2001}, and as a mathematical tool to construct a counterexample of additivity conjecture~\cite{H2009}.
Random states similarly have many utilities such as for saturating a classical communication capacity of noisy quantum channels~\cite{L1997}, 
for efficient measurements~\cite{RBSC2004,RRS2005,S2006} and for the estimation of gate fidelities~\cite{DCEL2009}.
Random states have been also often used to reveal generic properties of quantum states.
Since random states are uniformly distributed in a Hilbert space, their properties are supposed to show generic properties of quantum states.
For instance, entanglement of random states, called {\it generic entanglement}, has been intensively investigated from this point of view~\cite{L1978,P1993,FK1994,R1995,ZS2001,HLW2006,G2007,FMPPS2008,PFPFS2010,NMV2010,NMV2011}, and
it has been shown that almost all random states are almost maximally entangled.\\

In analogy with random unitary matrices and random states, {\it random diagonal-unitary matrices} and {\it phase-random states} have been proposed in Refs.~\cite{NTM2012,NM2013}.
They are originally introduced to investigate a typical behavior of time-independent Hamiltonian dynamics, but
are turned out to be useful for investigating typical properties of diagonal quantum circuits and of the states generated by them.
Random diagonal-unitary matrices also have applications in quantum informational tasks as we will see in Sects.~\ref{ssec:Gt} and~\ref{ssec:Therm}.

\begin{Definition}[{\bf Random diagonal-unitary matrices~\cite{NM2013} and Phase-random states~\cite{NTM2012}}]
{\it Random diagonal-unitary matrices in an orthonormal basis $\{ \ket{u_n} \}$, denoted by $\mathcal{U}_{\rm diag}(\{ \ket{u_n} \})$, are an ensemble of diagonal unitary matrices of the form $U_{\varphi} =  \sum_{n=1}^{d} e^{i \varphi_n} \ketbra{u_n}{u_n}$, where the phases $\varphi_n$ are uniformly distributed according to the normalized Lebesgue measure d$\varphi$ = d$\varphi_1 \cdots $d$\varphi_{d} / (2\pi)^{d} $ on $[0,2\pi)^d$.
Phase-random states are an ensemble of states $\{ U \ket{\Psi} \}_{U \in \mathcal{U}_{\rm diag}(\{ \ket{u_n} \})}$, which depends on the choice of the initial state $\ket{\Psi}$.}
\end{Definition}

As mentioned, we consider only random diagonal-unitary matrices and phase-random states in the computational basis in this paper.
Note that phase-random states depend on the choice of the initial state since random diagonal-unitary matrices do not have unitary invariance.
By choosing the initial state to be $\ket{+}^{\otimes n}$, the corresponding phase-random states are identified with an ensemble of all states that can be generated by IQP circuits.
Hence, generic properties of the states during the computation by IQP circuits can be revealed by investigating such phase-random states.
It has been shown that almost all phase-random states of which initial state is $\ket{+}^{\otimes n}$ are almost maximally entangled and they have even higher entanglement than generic entanglement of random states~\cite{NTM2012}. This implies that computation by IQP circuits possibly utilizes highly entangled states during the computation.

\subsubsection{$t$-designs} \label{ssec:t}

Although random (diagonal-)unitary matrices and (phase-)random states have been studied in quantum information from many perspectives, 
they cannot be efficiently generated since the number of parameters scales exponentially with the number of qubits.
Hence, it is important to introduce approximate ones, which are called $t$-designs.
In the following, we denote by $\mathbb{E}$ expectations over a probability distribution for simplicity.
If necessary, we specify the probability space taken over for the expectation.

A $t$-design of an ensemble is defined by an ensemble that simulates up to the $t$th-order statistical moments of the original ensemble on average,
and an $\epsilon$-approximate $t$-design is an ensemble that approximates the $t$-design, where $\epsilon$ is a degree of approximation.
In the case of approximate designs of matrices, the degree of approximation is often evaluated by the diamond norm~\cite{KSV2002}.
For a superoperator $\mathcal{E}$ acting on the bounded operators on a Hilbert space $\mathcal{H}$, the diamond norm is defined by
\begin{equation}
|\!| \mathcal{E} |\!|_{\diamond} := \sup_d \sup_{X \neq 0} \frac{|\!| (\mathcal{E} \otimes {\rm id}_d) X |\!|_1 }{ |\!|X|\!|_1}, \notag
\end{equation}
where ${\rm id}_d$ is the identity operator on another $d$-dimensional Hilbert space $\mathcal{H}^\prime$ and $X$ is any positive operator on $\mathcal{H} \otimes \mathcal{H^\prime}$.
To define an $\epsilon$-approximate $t$-design, let $\mathcal{V}$ be an ensemble of unitary matrices and
$\mathcal{E}_{\mathcal{V}}(\rho)$ be a superoperator such that 
\begin{align}
\mathcal{E}_{\mathcal{V}}^{(t)}(\rho) &:= \mathbb{E}_{U \in \mathcal{V}} [ U^{\otimes t} \rho (U^{\dagger})^{\otimes t}], \label{Eq:superop}
\end{align}
for any states $\rho$.
Then, an $\epsilon$-approximate unitary $t$-design is defined as follows (see, e.g., Ref.~\cite{HL2009}):

\begin{Definition}[{\bf $\epsilon$-approximate unitary $t$-designs~\cite{DCEL2009,TG2007}}] \label{Def:Ut}
{\it Let $\mathcal{U}$ be random unitary or diagonal-unitary matrices.
An $\epsilon$-approximate $t$-design of $\mathcal{U}$, denoted by $\mathcal{U}^{(t,\epsilon)}$, is an ensemble of unitary matrices such that
\begin{equation*}
|\!|  \mathcal{E}^{(t)}_{\mathcal{U}} - \mathcal{E}^{(t)}_{\mathcal{U}^{(t,\epsilon)}}  |\! |_{\diamond} \leq \epsilon.
\end{equation*}
The $t$-designs for random unitary and diagonal-unitary matrices are called {\it unitary} and {\it diagonal-unitary} $t$-designs, respectively. 
When $\epsilon=0$, the design is called exact.}
\end{Definition}

From a property of the diamond norm, an operational meaning of a $t$-design is given by that a unitary $t$-design cannot be distinguished from random unitary matrices even if we have $t$-copies of the system and apply any operations on them.
We note that
there are several definitions of an $\epsilon$-approximate unitary designs in terms of different measures of the distance\footnote{In several definitions of a $t$-design, it is required that the ensemble is finite, but we do not impose such a condition in this paper to be more general.}. 
However, they are all equivalent in the sense that,
if $\mathcal{V}$ is an $\epsilon$-approximate unitary $t$-design in one of the definitions,
then it is also an $\epsilon'$-approximate unitary $t$-design in other definitions, where $\epsilon' = {\rm poly} (2^{tn}) \epsilon$ (see Ref.~\cite{L2010} for details).

Similarly, an $\epsilon$-approximate state $t$-design is defined as follows.

\begin{Definition}[{\bf $\epsilon$-approximate state $t$-designs~\cite{RBSC2004,AE2007}}] \label{Def:appstate}
{\it Let $\Upsilon$ be random states or phase-random states.
An $\epsilon$-approximate $t$-design of $\Upsilon$, denoted by $\Upsilon^{(t,\epsilon)}$, is an ensemble of states such that
\begin{equation}
\biggl|\!\biggl|
\mathbb{E}_{\ket{\psi} \in \Upsilon^{(t,\epsilon)}} [ \ketbra{\psi}{\psi}^{\otimes t} ] 
- 
\mathbb{E}_{\ket{\psi} \in \Upsilon} [\ketbra{\psi}{\psi}^{\otimes t}]
 \biggr|\! \biggr|_{1} \leq \epsilon,
\end{equation}
where $|\!| \cdot |\!|_{1} = \tr | \cdot |$ is the trace norm.
In particular, we refer to a $t$-design of random states and phase-random states as a state $t$-design and a toric $t$-design, respectively.
When $\epsilon=0$, the design is called exact.}
\end{Definition}

In most applications of random unitary matrices and random states, their $t$-designs for small $t$ can be exploited~\cite{L2009}.
Hence, an efficient implementation of a unitary $t$-design, which also provides an efficient generation of a state $t$-design, is important for their applications.
In particular, an implementation of a unitary $t$-design by a {\it random circuit} has been intensely studied~\cite{ODP2007,DOP2007,Z2008,HL2009,DJ2011,BHH2012},
where a random circuit is a quantum circuit comprising random two-qubit gates that act on randomly chosen pairs of qubits.
In Ref.~\cite{BHH2012}, it has been shown that random circuits with a constraint that each two-qubit gate acts on nearest neighbor qubits
achieves an $\epsilon$-approximate $t$-design by applying $O(Nt^4(N+\log1/\epsilon))$ gates.

In contrast to the implementation of a unitary $t$-design, 
diagonal-unitary $t$-designs for general $t$ cannot be achieved by only one- and two-qubit {\it diagonal} gates.
This is because of the abelian property of the corresponding group
and a multi-qubit diagonal gate cannot be generally expressed by a product of diagonal matrices acting on smaller number of qubits.
For instance, a diagonal matrix ${\rm diag} (1,1,1,1,1,1,1,-1 )$ acting on three qubits cannot be decomposed into a product of diagonal two-qubit matrices.
Hence, implementing a diagonal unitary $t$-design by diagonal quantum circuits is a non-trivial task.
We review the results on the implementation of a diagonal-unitary $t$-design in the next section.

\subsection{Achieving a diagonal-unitary $t$-design by diagonal quantum circuits} \label{sec:DUbyIQP}

In this Section, we provide an efficient implementation of a diagonal-unitary $t$-design by diagonal quantum circuits.
This guarantees that the applications of a diagonal-unitary $t$-design, which will be given in Sects.~\ref{ssec:Gt} and~\ref{ssec:Therm}, are efficiently realizable by diagonal quantum circuits.
In particular, we consider to implement a diagonal-unitary $t$-design by a certain type of diagonal quantum circuits, called an {\it $r$-qubit phase-random circuit with a gate set $\mathcal{G}$} that generally containes multi-qubit gates. We introduce it in Sect.~\ref{ssec:PRC}.
In Sect.~\ref{ssec:ExactCons}, we provide a necessary and sufficient condition for an $r$-qubit phase-random circuit to achieve an exact diagonal-unitary $t$-design obtained in Ref.~\cite{NKM2013}.
In Sect.~\ref{ssec:ApproxCons},
we also show an approximate implementation of a diagonal-unitary $2$-design by using a simpler gate set composed only of the controlled-$Z$ gate and 
single-qubit random diagonal gates~\cite{NM2013}.
Since experimental manipulations of multi-qubit gates are not easy, this may help experimental implementations of the design.

\subsubsection{Phase-random circuit} \label{ssec:PRC}

\begin{figure}[tb!]
\centering
\begin{tabular}{ccc}
\begin{minipage}{0.33\hsize}
\centering
\includegraphics[width=40mm, clip]{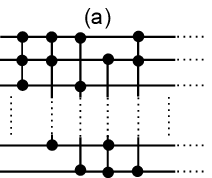}
\end{minipage}
\begin{minipage}{0.33\hsize}
\centering
\includegraphics[width=50mm, clip]{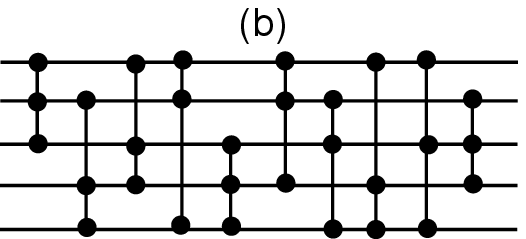} 
\end{minipage}
\begin{minipage}{0.33\hsize}
\centering
\includegraphics[width=50mm, clip]{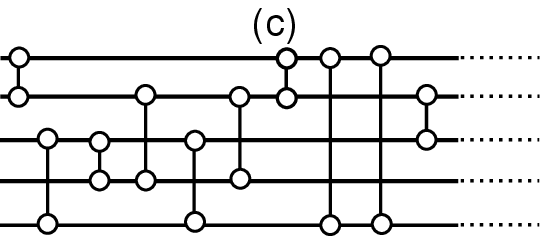}
\end{minipage}
\end{tabular}
  \caption{ 
  Panel (a) shows an $r$-qubit phase-random circuit with a diagonal gate set $\mathcal{G}_r$ in an $n$-qubit system, where $r=3$.
The black circles indicate the places of qubits $I_r$ on which the $r$-qubit gate acts on, and each $r$-qubit gate is randomly drawn from a diagonal gate set $\mathcal{G}_r$.
Panel (b) shows a $3$-qubit phase-random circuit used in Sect.~\ref{ssec:ExactCons} in a $5$-qubit system. 
In this case, the gates are applied on all combinations of $r$ qubits out of the $n$ qubits, so that $I_r$ is deterministically chosen.
Note that the order of the application is arbitrary since all gates are commutable.
Panel (c) shows the phase-random circuit with a gate set $\mathcal{G}_{\rm CZ}$ in Sect.~\ref{ssec:ApproxCons}.
The circle represents a single-qubit random diagonal gates followed by the controlled-$Z$ gate. The places of qubits $I_2$ are randomly chosen from $\{1,\cdots, n\}$.
}
\label{Fig:rPRC}
\end{figure}

An {\it $r$-qubit phase-random circuit with a gate set $\mathcal{G}_r$} consists of $T$ diagonal unitary gates. Each of the gates act on $r$ qubits.
For the $p$th gate, we choose $r$ different numbers $I_r^{(p)} \subset \{1,2,\cdots, n \}$ and apply an $r$-qubit diagonal gate $W_{I_r^{(p)}}$ on qubits
at sites $I_r^{(p)}$, where the gate $W_{I_r^{(p)}}$ is randomly chosen from a gate set $\mathcal{G}_r$.
An instance of the circuit is then specified by a set of parameters, $\mathcal{C}_T:=\{I_r^{(p)},W_{I_r^{(p)}}\}_{p=1}^T$, and the unitary operation corresponding to the circuit is given by $U_T = W_{I_r^{(T)}} W_{I_r^{(T-1)}} \cdots W_{I_r^{(1)}}$.
Thus, an $r$-qubit phase-random circuit with length $T$ is described by a set of the unitary operations $\{U_T\}_{\mathcal{C}_T}$.
See also Fig.~\ref{Fig:rPRC} (a).

\subsubsection{Exact implementation} \label{ssec:ExactCons}

Consider an $r$-qubit phase-random circuit with a diagonal gate set $\mathcal{G}_r$, where the gate set is given by
a set of diagonal gates with random phases in the computational basis,
\begin{equation}
\mathcal{G}_r = \{ {\rm diag} (e^{i \varphi_1}, e^{i \varphi_2}, \cdots, e^{i \varphi_{2^r}})\}_{\varphi_k \in [0,2 \pi)}. \label{Eq:G_r}
\end{equation}
We apply an $r$-qubit diagonal gates randomly drawn from the gate set $\mathcal{G}_r$ on all combinations of $r$ qubits out of $n$ qubits.
In this circuit, the choice of the place of qubits acted by the $p$th gate, $I_r^{(p)}$, is deterministic (see also Fig.~\ref{Fig:rPRC} (b)).
The random parameters in the circuit are then only phases in the diagonal gates,
leading to the probability measure on the phase-random circuit given by $\prod_{t=1}^T \prod_{k=1}^{2^r} d\varphi_k^{(t)}/2 \pi$.
The following Theorem provides a relation between $r$ and $t$ for this phase-random circuit to achieve a diagonal-unitary $t$-design.

\begin{Theorem}[{\bf Exact implementation of a diagonal-unitary $t$-design by Y. Nakata et al~\cite{NKM2013}}] \label{Thm:NecSuf_r}
{\it The $r$-qubit phase-random circuit with the gate set $\mathcal{G}_r$ defined above is an exact diagonal-unitary $t$-design if and only if
$r > \log_2 t $ for $t \leq 2^{n}-1$ and $r=n$ for $t \geq 2^N$.}
\end{Theorem}

Theorem~\ref{Thm:NecSuf_r} is obtained based on another equivalent definition of a $t$-design in terms of $\mathbb{E} [U^{\otimes t} \otimes (U^{\dagger})^{\otimes t}]$
(see e.g. Ref.~\cite{L2010}).
Then, we use a fact that
a matrix element in $U^{\otimes t} \otimes (U^{\dagger})^{\otimes t}$ containing a term $e^{i \varphi}$ becomes zero by averaging it over all $U \in \mathcal{C}_T$
since $\varphi$ is randomly chosen from $[0, 2 \pi)$.
By comparing the place of the constant terms in $U^{\otimes t} \otimes (U^{\dagger})^{\otimes t}$ for the phase-random circuit $\{U_T\}_{\mathcal{C}_T}$ and that for
random diagonal-unitary matrices $\mathcal{U}_{\rm diag}$, the statement of Theorem is reduced to a set identification problem.
The set identification problem can be solved in a combinatorial manner, and we obtain Theorem~\ref{Thm:NecSuf_r}.

The number of gates in the circuit is $\binom{n}{r}$, which should be $poly(n)$ for the circuit to have an efficient classical description, 
implying that $r$ should be constant.
Hence, Theorem~\ref{Thm:NecSuf_r} implies that a diagonal-unitary $t$-design can be efficiently implemented by the phase-random circuit when 
$t$ is constant with respect to the number of qubits.
If we restrict the circuit to use only two-qubit gates, we obtain an exact $t$-design only for $t\leq 3$.
This is contrasted to an implementation of an approximate unitary $t$-design for any $t$ by using two-qubit gates~\cite{BHH2012}.
This difference comes from a fact that there exist universal gate sets for non-diagonal quantum circuits, 
but does not exist a counterpart for diagonal quantum circuits, except a trivial one, if the gate set is restricted to be diagonal ones.
We also note that the gate set $\mathcal{G}_r$ can be replaced with a finite set of simpler multi-qubit gates. For the details, see Ref.~\cite{NKM2013}.

\subsubsection{Approximate implementation by a two-qubit gate set} \label{ssec:ApproxCons}

In Theorem~\ref{Thm:NecSuf_r}, the gate set $\mathcal{G}_r$ is chosen to be $r$-qubit diagonal gates with random phases. 
Since it may not be experimentally feasible to manipulate multi-qubit gates with random parameters,
it will be helpful to investigate what can be achieved by a fixed multi-qubit gate and random single-qubit diagonal gates.
Motivated by this, it was shown that a simpler gate set containing the controlled-$Z$ gate can achieve an approximate diagonal-unitary $2$-design~\cite{NM2013}.

Let us consider 
a $2$-qubit phase-random circuit with a diagonal gate set $\mathcal{G}_{\rm CZ}$ given by
\begin{equation}
\mathcal{G}_{\rm CZ} = \biggl\{\begin{pmatrix} 1 & 0 \\ 0 & e^{i \alpha} \end{pmatrix} \otimes \begin{pmatrix} 1 & 0 \\ 0 & e^{i \beta} \end{pmatrix}
\begin{pmatrix} 1 & 0 & 0 & 0 \\ 0 & 1 & 0 & 0 \\ 0 & 0 & 1 & 0 \\ 0 & 0 & 0 & -1 \end{pmatrix} \biggr\}_{ \alpha, \beta \in [ 0, 2 \pi)}.
\end{equation}
In this case, we choose a pair of two qubits $I_2^{(p)}$ for the $p$-th gate randomly from $\{1,\cdots, n\}$ (see Fig.~\ref{Fig:rPRC} (c)).
Thus, the probability measure of the circuit is given by $\prod_{p=1}^T \frac{2}{N(N-1)} d \alpha_p  d\beta_p/(2\pi)^2$.
In this phase-random circuit, there exist terms in 
$\mathbb{E}_{\mathcal{C}_T}[U^{\otimes t} \otimes (U^{\dagger})^{\otimes t}]$ that are equal to $-1$ due to the use of the controlled-$Z$ gate,
while all elements in $\mathbb{E}_{\mathcal{U}_{\rm diag}}[U^{\otimes t} \otimes (U^{\dagger})^{\otimes t}]$ are either $0$ or $1$.
Consequently, the phase-random circuit does not achieve an exact diagonal-unitary $t$-design.
Nevertheless, it achieves an $\epsilon$-approximate diagonal-unitary $2$-design by applying at most
$O(N^2(N+\log \epsilon^{-1}))$ gates randomly drawn from the gate set $\mathcal{G}_{\rm CZ}$ as stated in the following Theorem.

\begin{Theorem}[{\bf Approximate implementation of a diagonal 2-design by Y. Nakata et al~\cite{NM2013}}] \label{Thm:phase-random circuitCZ}
{\it The $2$-qubit phase-random circuit with a gate set $\mathcal{G}_{\rm CZ}$ of a length $T$ is an $\epsilon$-approximate diagonal-unitary $2$-design 
if $T \geq T_{conv}(\epsilon)$, where 
\begin{equation}
\frac{N}{2}  + \biggl(\frac{N^2}{4} + O(N) \biggr)\log (2\epsilon)^{-1}  \leq T_{conv}(\epsilon) 
\leq 7 N^3 \log2 + N^2 \log\epsilon^{-1} + O(N^2).  
\end{equation}
Therefore, the phase-random circuit is an $\epsilon$-approximate diagonal-unitary $2$-design after applying at most
$O(N^2(N+\log \epsilon^{-1}))$ two-qubit gates.}
\end{Theorem}

Theorem~\ref{Thm:phase-random circuitCZ} is proven by a method developed in Ref.~\cite{ODP2007,DOP2007}, which maps the transformation of a state by the circuit 
to a Markov chain on a certain graph.
By investigating the mixing time of the Markov chain, we obtain Theorem~\ref{Thm:phase-random circuitCZ}.
Note that a random choice of a pair of two qubits in the phase-random circuit is crucial in Theorem~\ref{Thm:phase-random circuitCZ}.
If the qubits are deterministically chosen, the commutativity of gates makes the circuit stationary after the two-qubit gates are applied on all pairs of qubits.
It is easy to check that such a stationary circuit is not a diagonal-unitary $2$-design.
When the two qubits are randomly chosen, the classical randomness prevents the circuit from being stationary even after 
the gates are applied on all pairs of qubits. 
This is also clear from the probability measure of the phase-random circuit.
As a result, the degree of approximation is reduced to be arbitrarily small and obtain 
an $\epsilon$-approximate diagonal-unitary $2$-design.

\subsection{Generating a state $t$-design by a diagonal-unitary $t$-design} \label{ssec:Gt}

In Ref.~\cite{NKM2013}, it was shown that applying a diagonal-unitary $t$-design on a specific initial state achieves a good approximate {\it state} $t$-design
for any $t$, although the degree of approximation is constant.
The degree of approximation can be improved by combining a diagonal-unitary $t$-design with a classical random procedure,
and particularly, an exact state $2$-design are obtained~\cite{NM2013}.
Since a state $t$-design for small $t$ is used in many quantum informational tasks and 
diagonal-unitary $t$-design can be implemented by diagonal quantum circuits, these results provides an application of diagonal quantum circuits.
We review a generation of an approximate state $t$-design in Sect.~\ref{sssec:Approx}, and an exact one for $t=2$ in Sect.~\ref{sssec:Exact}.

\subsubsection{Approximate state $t$-design} \label{sssec:Approx}

The following proposition states that a diagonal-unitary $t$-design with an appropriate separable initial state 
generates an approximate state $t$-design.

\begin{Proposition}[{\bf Generating an approximate state $t$-design by Y. Nakata et al~\cite{NKM2013}}] \label{Prop:Protocol1}
{\it An ensemble of states obtained 
by applying a diagonal-unitary $t$-design to an initial state $\ket{+}^{\otimes n}$
is an $\eta(n,t)$-approximate state $t$-design, where 
\begin{equation}
\eta(N,t)=\frac{t(t-1)}{2^n} + O(\frac{1}{2^{2n}}).
\end{equation}}
\end{Proposition}

Although the degree of approximation $\eta(n,t)$ is constant and cannot be improved by applying additional diagonal gates, it is already a good approximation for $t$
independent of $n$.
This good approximation of random states by the phase-random states is a consequence of the {\it concentration of measure} of random states~\cite{L2001}.
The concentration of measure in this case means that
almost all random states are equal-amplitudes states in the sense that
$\ket{\Psi} = \sum_{k=1}^{2^n} c_k \ket{u_k}$, where $|c_k| \sim 2^{-n/2}$ and $\{ \ket{u_k} \}$ is some orthonormal basis.
Hence, random states can be well-approximated by an ensemble of states with equal-amplitudes in a fixed basis.
However, the distribution of amplitudes of states in an ensemble generally depends on the basis, 
which differs from the distribution of coefficients of random states independent of the basis due to the unitary invariance. 
This makes the ensemble in Proposition~\ref{Prop:Protocol1} not an exact but approximate state design.

Proposition~\ref{Prop:Protocol1} implies that a diagonal-unitary $t$-design is capable to generate an ensemble of states of which distribution is 
hard to distinguish from the uniform one as long as looking at lower order statistical moments.
Since a diagonal-unitary $t$-design is obtained by a phase-random circuit with an initial state $\ket{+}^{\otimes n}$, which is an IQP circuit,
computation by IQP circuits is typically exploiting uniformly distributing states in the Hilbert space, as mentioned in Sect.~\ref{ssec:IQP}.

\subsubsection{Exact state $2$-design by virtue of classical randomness} \label{sssec:Exact}

Although a diagonal quantum circuit achieves only an approximate state $t$-design with a fixed degree of approximation,
the degree of approximation can be improved by combining them with a classical random procedure.
In particular, the resulting ensemble becomes an exact design when $t=2$ as stated below. 

\begin{Proposition}[{\bf Generating an exact state $2$-design by Y. Nakata et al~\cite{NM2013}}] \label{Prop:2exact}
{\it Consider the following protocol.
\begin{enumerate}
\item With probability $\frac{1}{2^n+1}$, choose a random $n$-bit sequence $\bar{m}$ and generate a state $\ket{\bar{m}}$.
\item With probability $\frac{2^n}{2^n+1}$, apply a diagonal-unitary $2$-design on an initial state  $\ket{\vec{+}}=\ket{++ \cdots +}$.
\end{enumerate}
Then, the resulting ensemble is an exact state $2$-design.}
\end{Proposition}

Proposition~\ref{Prop:2exact} is simply obtained by looking at the difference between
$\mathbb{E}_{\ket{\psi} \in \Upsilon^{(2)}_{\rm Haar}} [\ketbra{\psi}{\psi}^{\otimes 2}]$, where $ \Upsilon^{(2)}_{\rm Haar}$ represents a state $2$-design, and
$\mathbb{E}_{U \in \mathcal{U}^{(2)}_{\rm diag}} [( U \ketbra{\vec{+}}{\vec{+}} U^{\dagger})^{\otimes 2}]$, which is given by
\begin{align}
\mathbb{E}_{\ket{\psi} \in \Upsilon^{(2)}_{\rm Haar}} [\ketbra{\psi}{\psi}^{\otimes 2}]
=
\frac{2^n}{2^n +1}
\mathbb{E}_{U \in \mathcal{U}^{(2)}_{\rm diag}} [( U \ketbra{\vec{+}}{\vec{+}} U^{\dagger})^{\otimes 2}]
+
\frac{1}{2^n + 1}\sum_{m} \ketbra{\bar{m}}{\bar{m}}^{\otimes 2}. \label{Eq:protexact}
\end{align}
The equation~\eqref{Eq:protexact} implies that an exact state $2$-design is obtained by a probabilistic mixture of the ensemble generated by a diagonal-unitary $2$-design 
$\{ U \ketbra{\vec{+}}{\vec{+}} U^{\dagger}\}_{U \in \mathcal{U}^{(2)}_{\rm diag}}$ and 
product states $\{ \ket{\bar{m}} \}$.

This protocol of generating an exact state $2$-design has experimental merits compared to previously known protocols listed below:

\begin{itemize}
\item A generation of an exact unitary 2-design using Clifford operations is known~\cite{DLT2002}, which requires $O(n^8)$ bits and $O(n^2)$ quantum gates. 
In the protocol, unitary matrices of generating an exact design are classically calculated and are decomposed into one- and two-qubit unitary gates. 
Thus, for obtaining a state $2$-design,
it needs to repeat calculating a gate decomposition and constructing the corresponding quantum circuit.
\item An $\epsilon$-approximate unitary 2-design is obtained by a quantum circuit composed of one- and two-qubit Clifford gates, where 
some gates are applied probabilistically~\cite{DCEL2009}. The number of gates is $O(n(n+ \log 1/\epsilon))$~\cite{HL2009} in Definition~\ref{Def:appstate}.
\item 
A random circuit~\cite{ODP2007,DOP2007} with length $O(n(n+ \log 1/\epsilon))$ is an $\epsilon$-approximate unitary 2-design~\cite{HL2009,DJ2011},
where all gates are randomly chosen from a gate set called a {\it 2-copy gapped} gate set, e.g., a set of the controlled-NOT gate and random single-qubit gates.
\item A local random circuit with length $O(n t^4( n + \log 1/\epsilon))$ gates is an $\epsilon$-approximate unitary $t$-design~\cite{BHH2012} for any $t$. 
The circuit is composed of random $SU(4)$ gates acting on nearest neighbor qubits.
\end{itemize}

The protocol in Proposition~\ref{Prop:2exact} is experimentally preferable because of the following reasons.
First, a diagonal-unitary $2$-design is achieved by diagonal quantum circuits comprising only two-qubit diagonal gates.
Moreover, 
the circuits are implementable by a single time-independent commuting Hamiltonian due to the commutativity of all the gates.
Since such a Hamiltonian can be simultaneously applied on all qubits, the practical time of an implementation is significantly reduced compared to non-diagonal ones.
Hence, the implementation of a quantum part in the protocol is easier and more robust than other protocols.
Finally, most of the previous protocols except the first one achieve only an {\it approximate} $2$-design,
while the protocol in Proposition~\ref{Prop:2exact} achieves an {\it exact} one.
We also emphasize that there is no drawback in the protocol of Proposition~\ref{Prop:2exact} since the same number of gates as that of previous ones are used in the protocol.

One may wonder how much the degree of approximation is improved by adding classical randomness in the case of general $t$.
It was shown that the improvement is limited to be $O(2^{(1-t)n})$ for general $t$~\cite{NKM2013}.
Since the degree of approximation without classical randomness is $\eta(n,t) = O(1/2^n)$,
the protocol does not result in a significant improvement except for $t=2$.

\subsection{Verifying the principle of apparently equal a priori probability in quantum statistical mechanics} \label{ssec:Therm}

Diagonal quantum circuits can be also used to verify the foundation of quantum statistical mechanics.
In quantum statistical mechanics, a derivation of a canonical thermal state $e^{-\beta H}/\tr e^{-\beta H}$ for a given Hamiltonian $H$ with an inverse temperature $\beta$
from natural assumptions is one of the fundamental problems~\cite{N1955,S1989,H1998,GLTZ2006}. 
Recently, a new development on the problem based on {\it the principle of apparently equal a priori probability} has been made~\cite{PSW2006}.
We introduce a quantum algorithm of generating a thermal state based on the principle by using a diagonal quantum circuit, which we call a {\it thermalizing algorithm}.
We overview the principle in Sect.~\ref{sssec:can} and provide the thermalizing algorithm for a certain class of Hamiltonians in Sect.~\ref{sssec:themal}.

\subsubsection{Appearance of a canonical thermal state} \label{sssec:can}

In Ref.~\cite{PSW2006}, it was shown that a standard assumption for the derivation of canonical thermal states, i.e., {\it the equal a priori probability postulate} which is also known as a {\it microcanonical assumption}, can be replaced by a weaker assumption.
To clarify the situation, let us consider a composite system $S+B$, where $S$ and $B$ represent a system and a thermal bath, respectively,
and denote their Hilbert spaces by $\mathcal{H}_S$ and $\mathcal{H}_B$, respectively.
Let $H_{\rm tot}=H_S+H_B+H_{\rm int}$ be a Hamiltonian acting on $\mathcal{H}_S \otimes \mathcal{H}_B$, where $H_S$ ($H_B$) acts on only $\mathcal{H}_S$ ($\mathcal{H}_B$) and
$H_{\rm int}$ acts on both. 
We denote the eigen decomposition of $H_{\rm tot}$ by $\sum_i e_i \ketbra{e_i}{e_i}$ and define a subspace restricted by total energy $E$, 
$\mathcal{H}_E = {\rm span}\{ \ket{e_i} | E - \delta < e_i< E\}$, where $\delta$ is supposed to be sufficiently small.
Then, the equal a priori probability postulate states that 
it is always the case in the systems describable by thermodynamics that the equiprobable state $\mathbb{I}_{\mathcal{H}_E}/\tr \mathbb{I}_{\mathcal{H}_E}$ is realized, where $\mathbb{I}_{\mathcal{H}_E}$ is a projector onto $\mathcal{H}_E$.
This postulate leads to a thermal state in the system $S$ in the thermodynamic limit, i.e., $\tr_B \mathbb{I}_{\mathcal{H}_E}/\tr \mathbb{I}_{\mathcal{H}_E} = 
e^{-\beta H_S}/\tr e^{-\beta H_S}$ where the inverse temperature $\beta$ is determined by the energy $E$ in the total system $S+B$.
However, the assumption is very strong since it is a statement about one single initial state.

What has shown in Ref.~\cite{PSW2006} is that the assumption of the equiprobable state can be relaxed to random states in $\mathcal{H}_E$.
This is the principle of apparently equal a priori probability.
More precisely, it has been shown that, for almost all random states $\ket{\Psi}$ in $\mathcal{H}_E$, the reduced density state in the system $S$ is very close to 
that of the equiprobable state, $\tr_{B} \ketbra{\Psi}{\Psi} \sim \tr_B \mathbb{I}_{\mathcal{H}_E}/\tr \mathbb{I}_{\mathcal{H}_E} = e^{-\beta H_S}/\tr e^{-\beta H_S}$.
Since this implies that a state randomly drawn from the Hilbert space $\mathcal{H}_E$ results in a thermal state in a system $S$ with high probability,
it is a stronger statement than the equal a priori probability postulate.
Similarly, it has been shown that this is also the case for certain types of phase-random states~\cite{NTM2012},
that is, almost all phase-random states in the subspace $\mathcal{H}_E$ locally equilibrate to a thermal state in the above sense if the basis and the initial state of the phase-random states satisfy certain conditions.
In particular, if the initial state is an equal superposition of all eigenstates in $\mathcal{H}_E$, the condition is satisfied.

\subsubsection{A thermalizing algorithm} \label{sssec:themal}

		\begin{figure}[tb]
		\centering
		\includegraphics[width=70mm, clip]{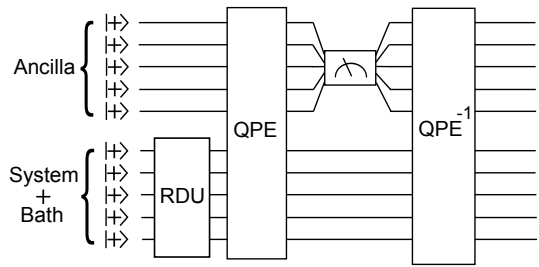}
		\caption{A quantum circuit for implementing thermalizing algorithm of classical Hamiltonians by using random diagonal-unitary matrices. The upper half of the circuit represents ancilla qubits and the lower half represents the system and the artificial thermal bath.
RDU and QPE (QPE$^{-1}$) denote random diagonal-unitary matrices and the (inverse of) quantum phase estimation, respectively. }
		\label{Fig:THM}
		\end{figure}

The principle of apparently equal a priori probability offers a possibility of a quantum algorithm that realizes a thermal state by using random and phase-random states in $\mathcal{H}_E$.
Here, we particularly provide a quantum algorithm using a diagonal quantum circuit, which efficiently realizes a thermal state of classical Hamiltonians.
By classical Hamiltonians, we mean those with separable eigenstates.
Such algorithms are important from the viewpoint not only of an experimental verification of the foundation of quantum statistical mechanics, 
but also of obtaining a thermal state of arbitrary classical Hamiltonian in experiments.
The latter has utilities in condensed matter physics, since it enables us to measure expectation values of any observables easily in experiments,
and also has a possible application in quantum information processing that exploits a thermal state~\cite{FNOM2013}.
Note that there are several algorithms that realize a thermal state of a given Hamiltonian without knowing its eigenenergies and eigenstates, e.g.,
quantum Metropolis algorithm~\cite{TOVPV2011} and an artificial thermalization circuit~\cite{RGE2012}, which are based on different mechanisms from the principle of apparently equal a priori probability.

As shown in Ref.~\cite{NTM2012}, it is sufficient for realizing a thermal state to prepare phase-random states with equal-amplitudes in the computational basis in the subspace $\mathcal{H}_E$.
To see this more clearly, let us consider $n=n_S + n_B$ qubits, where $n_S$ and $n_B$ are the number of qubits in the system $S$ and in the thermal bath $B$, respectively. 
We denote by $\{ \ket{e_l^E} \}_{l=1,\cdots, d_E}$ the eigenstates of $H_{\rm tot}$ in the restricted subspace $\mathcal{H}_E$.
Note that they are a subset of the computational basis by the assumption that the Hamiltonian is classical.
The principle of apparently equal a priori probability based on phase-random states in $\mathcal{H}_E$~\cite{NTM2012} states that 
almost all phase-random states in the form of
\begin{equation}
\frac{1}{\sqrt{d_E}} \sum_{l=1}^{d_E} e^{i \varphi_l} \ket{e_l^E},
\end{equation}
are locally close to a thermal state.
This state is approximately achieved by the quantum circuit presented in Fig.~\ref{Fig:THM}.

The circuit uses $r$ ancilla qubits initially prepared in $\ket{+}^{\otimes r}$, random diagonal-unitary matrices, the quantum phase estimation (QPE)~\cite{K1995,NC2000} and the projective measurement on the ancilla qubits in the computational basis. 
Random diagonal-unitary matrices can be replaced by a diagonal-unitary $t$-design by a similar argument in Ref.~\cite{L2009},
and the most parts of QPE except the quantum Fourier transformation can be also implemented by a diagonal quantum circuit since the Hamiltonians are classical.
At each part of the circuit, the state changes as follows:
\begin{enumerate}
\item The random diagonal-unitary matrices acting on the system and the thermal bath add random phases $\{\varphi_l\}$ to $\ket{+}^{\otimes n}$,
and generates a phase-random state $2^{-n/2} \sum_{l=1}^{2^n} e^{i \varphi_l} \ket{\bar{l}}$ in $\mathcal{H}_S+\mathcal{H}_B$.
\item After QPE, the state is approximately $\ket{\Psi_{e}}=2^{-n/2} \sum e^{i \varphi_l} \ket{\bar{l}} \otimes \ket{\bar{e}_l}$ where $\bar{e}_l$ is a binary representation of the eigenenergy $e_l$ of the total Hamiltonian $H_{\rm tot}$.
\item By performing the projective measurement $P:=\{ P_E, P_{\neg E}  \}$, where $P_E$ is a projection operator onto $\mathcal{H}_E$ and $P_{\neg E} = I-P_E$, on the ancilla qubits, the state is probabilistically changed into $\frac{1}{\sqrt{d_E}} \sum_{l=1}^{d_E} e^{i \varphi_l} \ket{e_l^E} \otimes \ket{\bar{e}_l}$.
Note that this measurement is done in the computational basis.
The success probability is given by a probability to obtain the outcome corresponding to $P_E$.
Otherwise, the algorithm fails.
\item The inverse of QPE changes the state back to $\ket{\Psi_f}= (\frac{1}{\sqrt{d_E}} \sum_{l=1}^{d_E} e^{i \varphi_l} \ket{e_l^E}) \otimes \ket{+}^r$, 
so that we obtain a phase-random state in the subspace $\mathcal{H}_E$ by tracing out the ancilla qubits.
\end{enumerate}

There are two error factors in the algorithm, which originates from QPE.
First, the eigenenergy is approximated by binary numbers within a precision of $2^{-r}$. This approximation results in a round-off error~\cite{TOVPV2011}. 
Second, QPE does not exactly transform the state to $\ket{\Psi_{e}}$. This is inherited in the final state, resulting in $\sqrt{(1-\epsilon)}\ket{\Psi_f} + \sqrt{\epsilon} \ket{\Psi_{Error}}$. However, these errors can be sufficiently suppressed by preparing a large number of ancilla qubits such that $2^{-r} \ll \Delta e$, where $\Delta e$ is the minimum energy gap of $H_{\rm tot}$.  It is often the case for local Hamiltonians that $\Delta e$ scales at worst exponentially with the number of particles $n$.
Hence, the error of the algorithm is sufficiently small if $r$ is chosen to be $poly(n)$. 
Note that obtaining a thermal state at low temperatures is generally difficult since the probability to obtain $P_E$ in the projective measurement depends on the temperature as shown in Ref.~\cite{RGE2012}.

Although the main aim of our algorithm is an experimental verification of the principle of apparently equal a priori probability,
it has an advantage to some extent for the purpose of obtaining a thermal state even compared to other thermalizing algorithms~\cite{TOVPV2011,RGE2012} 
since most parts of the algorithm are implementable
by diagonal quantum circuits.
On the other hand, it has certain drawbacks.
One is that our algorithm needs to use the Hadamard gate in the quantum Fourier transformation in QPE. 
The number of the Hadamard gates is $r$, which is the number of ancilla qubits and determines the error of the algorithm.
Thus, there is a trade-off relation between the number of the Hadamard gates and the precision of the algorithm.
The other is that our algorithm works only for classical Hamiltonians. This drawback can be solved if random diagonal-unitary matrices in our algorithm are replaced 
by random unitary matrices. However, such an algorithm necessarily requires non-diagonal quantum circuits, which is probably not feasible by current experimental technology.

\section{Summary and concluding remarks} \label{sec:Sum}

In this paper, we have reviewed a study of diagonal quantum circuits in the computational basis
motivated by theoretical and practical interests.
From the theoretical point of view, diagonal quantum circuits are a good framework to investigate the origin of quantum speed-up 
since quantum computation described by diagonal quantum circuits is supposed to be on the boarder of classical and quantum computation.
Based on this idea, we have reviewed the computational power of IQP circuits in terms of classical simulatability.
It has been shown that the output probability distribution of IQP circuits is in general highly implausible to be classically simulated.
This is the case even if the circuits are composed only of $2$-qubit diagonal gates.
However, there also exist IQP circuits that are classically simulatable.


On the other hand, diagonal quantum circuits have a practical importance since realizations of diagonal gates in experiments are more feasible than non-diagonal gates.
They are implementable in a fault-tolerant manner by current technology, so that 
any quantum tasks using diagonal quantum circuits are likely to be already realizable in experiments.
We have reviewed two applications of diagonal quantum circuits, generating an approximate state $t$-design and a thermalizing algorithm.
Such applications enable us to experimentally demonstrate quantum advantages in informational tasks and quantum nature in statistical mechanics.

It is worth investigating the diagonal quantum circuits further from both theoretical and practical point of view.
On classical simulatability of IQP circuits, it is important to study the complete classification of IQP circuits
since 
clarifying computational power of IQP circuits will lead a better understanding of the origin of quantum speed-up.
Since IQP circuits have a simple structure,
this approach is more suitable than studying a distinction between quantum computation by non-diagonal quantum circuits and classical one.
It is also interesting to consider a difficulty of average instances of IQP circuits.
The results about hardness of classical simulation reviewed in this paper are obtained by deriving a highly implausible statement from 
an assumption that {\it all} IQP circuits can be classical simulatable.
This methodology, however, does not answer to the question about hardness of classical simulatability of a {\it specific} IQP circuit.
One way to address this question is to study classical simulatability of randomly chosen IQP circuits, which is an idea of the study of average instances in computational complexity theory.
Investigating a difficulty of average instances of IQP circuits is important to fully understand classical simulatability of IQP circuits.
It is also practically desirable to consider what diagonal quantum circuits can perform beyond classical information processing.
This will be an important step toward a realization of a quantum computer since such tasks are experimentally realizable to
demonstrate quantum advantages and will accelerate an experimental challenge of making a quantum computer.
One possible application of diagonal quantum circuits is a {\it decoupling} of two systems~\cite{DBWR2010,BF2013}, which is often used in quantum informational tasks.
The decoupling is originally proposed by using random unitary matrices, but it is not necessary to use them.
Since random diagonal-unitary matrices have similar properties to random unitary matrices in some aspects,
an approximate decoupling may be achievable by using diagonal-unitary designs realized by diagonal quantum circuits.\\

This work is supported by Project for Developing Innovation Systems of the Ministry of Education, Culture, Sports, Science and Technology (MEXT), Japan. 
Y. N. acknowledges JSPS Postdoctoral Fellowships for Research Abroad.
M.~M acknowledge support from JSPS by KAKENHI, Grant No. 23540463.
M. M also acknowledge to the ELC project (Grand No. 24106009) for encouraging the research project in this paper.

%
%
%

%
%

\end{document}